\documentstyle[preprint2,aasms4]{aastex}
\doublespace{\parskip 4pt plus 1.5pt
\baselineskip 26pt plus 1pt minus .5pt
\lineskip 6pt plus 1pt \lineskiplimit 5pt }
\def\beq{\begin{equation}}
\def\eeq{\end{equation}}
\def\beqar{\begin{eqnarray}}
\def\eeqar{\end{eqnarray}}
\newcommand{\B}{{\bf B}}

\newcommand{\ra}{{\rangle}}
\newcommand{\laa}{{\langle}}

\newcommand{\bu}{{\bf u}}
\newcommand{\bbv}{{\bf v}}
\newcommand{\dt}{{\delta t}}
\newcommand{\bb}{{\bf b}}
\newcommand{\curl}{{\nabla \times}}

\newcommand{\lap}{{\nabla^2 }}
\newcommand{\q}{{\bf q}}
\newcommand{\x}{{\bf x}}
\newcommand{\y}{{\bf y}}

\newcommand{\k}{{\bf k}}
\newcommand{\p}{{\partial}}

\newcommand{\aalpha}{{\xi_{v}}}
\newcommand{\bbeta}{{\xi_{b}}}
\newcommand{\gapprox}{\lower.4ex\hbox{$\;\buildrel
>\over{\scriptstyle\sim}\;$}}
\newcommand{\lapprox}{\lower.4ex\hbox{$\;\buildrel
<\over{\scriptstyle\sim}\;$}}

\lefthead{Kim \& Diamond}
\righthead{On turbulent reconnection}

\begin{document}
\def\ref#1{\hangindent=6em\hangafter=1 {#1}}

\title{ON TURBULENT RECONNECTION}
\author{EUN-JIN KIM and P. H. DIAMOND} 
\affil{Department of Physics, University of California
at San Diego, La Jolla, CA 92093, USA}

\begin{abstract}
We examine the dynamics of turbulent reconnection in
2D and 3D reduced MHD by calculating the effective
dissipation due to coupling between small--scale fluctuations
and large--scale magnetic fields. Sweet--Parker type
balance relations are then used to calculate the global
reconnection rate. Two approaches are employed --- quasi--linear
closure and an eddy-damped fluid model. Results indicate
that despite the presence of turbulence,
the reconnection rate remains inversely proportional
to $\sqrt{R_m}$, as in the Sweet--Parker analysis.
In 2D, the global reconnection rate is shown to be enhanced
over the Sweet--Parker result by a factor of magnetic Mach
number. These results are the consequences of the constraint
imposed on the global reconnection rate by the requirement
of mean square magnetic potential balance. The incompatibility
of turbulent fluid--magnetic energy equipartition and
stationarity of mean square magnetic potential is demonstrated. 

{\it Subject headings: MHD --- magnetic fields --- turbulence} 

\end{abstract}
\pagebreak

\section{INTRODUCTION}
Magnetic reconnection is the process whereby large scale magnetic
field energy is dissipated and magnetic topology is altered in 
MHD fluids and plasmas (for instance, see, Vasyliunas 1975; 
Parker 1979; Forbes \& Priest 1984; Biskamp 1993;
Wang, Ma, \& Bhattacharjee 1996 and references therein). 
Reconnection is often invoked as the
explanation of large scale magnetic energy release in space,
astrophysical, and laboratory plasmas. Specifically, magnetic 
reconnection is thought to play an integral role in the dynamics of
the magnetotail, the solar dynamo, solar coronal heating, and
in the major disruption  in tokamaks. For these reasons, magnetic
reconnection has been extensively studied in the context of
MHD, two-fluid and kinetic models, via theory,
numerical simulations and laboratory experiments.

The basic paradigm for magnetic reconnection is the Sweet--Parker
(called SP hereafter) problem (Parker 1957; Sweet 1958), 
in which a steady inflow velocity
advects oppositely directed magnetic field lines ($\pm {\bf B}$)
together, resulting in current sheet formation and, thus, reconnection
(see Fig. 1). The current sheet has thickness $\Delta$ and length
$L$, so that imposition of continuity ($v_r L = v_0 \Delta$),
momentum balance ($v_0 = v_A$) and magnetic energy balance
($v_r B = \eta B/\Delta$) constrains the inflow, or
``reconnection'', velocity to be 
$v_r = v_A/\sqrt{S} \propto v_A/\sqrt{R_m}$. 
Here $v_0$ is the outflow velocity; 
$v_A$ is the Alfv\'en speed associated with $\B$;
$S \equiv v_A L/\eta$ is the Lundquist number; 
$R_m = ul/\eta$ is the magnetic Reynolds number, with $u$ 
and $l$ being the characteristic amplitude and length scale of 
the velocity --- $S$ is called the magnetic Reynolds number 
$R_m$ in some literatures. 
Note that the SP process forms strangely anisotropic current sheets
since $\Delta/L = \sqrt{S}$ and $S \gg 1$. Note also the link
between sheet anisotropy and the reconnection speed $v_r$, i.e.,
$v_r/v_A = \Delta/L = 1/\sqrt{S}$.
Finally, it should be
noted that $v_r$ is a measure of the {\it global} reconnection
rate, in that it parameterizes the mean inflow velocity to the
layer. 

The SP Picture is intrinsically appealing, on account of its 
simplicity and dependence only upon conservation laws.
Moreover, the SP prediction has been verified by laboratory
experiments (Ji, Yamada, \& Kulsrud 1998). 
However, since $R_m$ is extremely
large in most astrophysical applications of interest 
(i.e. $R_m \sim 10^{13}$ in the solar corona), the SP
reconnection speed is pathetically slow. Hence, there
have been many attempts to develop models of {\it fast}
reconnection. For example, in 1964 Petschek proposed a
fast reconnection model involving shock formation near the
reconnection layer, which predicted $v_r = v_A/\ln{S}$.
Unfortunately, subsequent numerical (Biskamp 1986)
and theoretical (Kulsrud 2000) study has indicated that
Petschek's model is internally inconsistent. While
research on fast, laminar reconnection continues today (i.e.,
Kleva, Drake, \& Waelbroeck 1995) in the context of 
two-fluid models, the failure
of the Petschek scenario has sparked
increased interest in turbulent reconnection (Matthaeus \& Lamkin 1986)
in which turbulent transport coefficients (which can be
large for large Reynolds number) act as effective
dissipation coefficients, and so are thought to
facilitate fast reconnection (i.e., Diamond et al. 1984; Strauss 1988).
Interest in turbulent reconnection has also been stimulated by
the fact that many instances of reconnection occur in systems
where turbulence is ubiquitous, i.e., coronal heating
of turbulent accretion disks, the dynamo in the sun's
convection zone, and turbulent tokamak plasmas during
disruptions.

Recently, Lazarian and Vishniac (1999) (referred to hereafter
as LV) presented a detailed discussion of turbulent
reconnection. LV took a rather novel approach to the
problem by considering the interaction of two slabs of
oppositely directed, chaotic magnetic fields when
advected together. LV modeled the effects of
turbulence by treating the slabs' surfaces as
rough, where the roughness was symptomatic of a
chaotic turbulent magnetic field structure. This
`rough surface' model naturally led LV to decompose the
reconnection process into an ensemble of local,
`micro'-reconnection events, which interact to form a
net `global' reconnection process. LV argue that micro-reconnection
events occur in small scale `layers', with dimensions set by
the structure of the underlying Alfv\'enic MHD turbulence
(i.e., the $k_{\perp}^{-1}$ and $k_{\parallel}^{-1}$,
as set by the Goldreich--Sridhar model). The upper
bound for the micro-reconnection rate obtained by LV
is $v_r = v_A (u/v_A)^2 = v_A (b/B_H)^2$, where $B_H$
is the mean, reconnection field, and $u$ and $b$ are
small--scale velocity and magnetic field. While the LV arguments
concerning micro-reconnection are at least plausible,
their assertion that the global reconnection rate
can be obtained by effectively superposing micro--reconnection
events is unsubstantiated and rather dubious, in that it
neglects dynamical interactions between micro-layers.
Such interactions are particularly important for enforcing
topological conservation laws. Since the process of turbulent
reconnection is intimately related to the rate of flux
dissipation, and the latter is severely constrained by
mean square magnetic potential conservation, it stands
to reason that such a topological conservation law will also
constrain the rate of {\it global} magnetic reconnection.
In particular, for a mean $B$-field with strength in excess
of $B_{{\rm crit}} \sim \sqrt{\laa u^2 \ra/R_m}$, the flux
2D was shown to be suppressed by a factor 

\begin{equation}
{1 \over 1 + R_m {\langle B \rangle^2 / \langle u^2 \rangle }}\,,
\label{eq0}
\end{equation}
where $\langle B \rangle$ is the large--scale magnetic field 
and $\langle u^2 \rangle$ the turbulent kinetic energy
(Cattaneo \& Vainshtein 1991; Gruzinov \& Diamond 1994); 
The above expression implies that even a weak magnetic
field (i.e, one far below the equipartition value $\laa b^2 \ra
\sim \laa u^2 \ra$) is potentially important. The origin
of this suppression is ultimately linked to the conservation
of mean square potential (see Das \& Diamond 2000 for flux
diffusion in EMHD). Hence, it is natural to investigate
the effect of such constraints on reconnection, as well.

In turbulent reconnection, fluctuating magnetic fields are
dynamically coupled to a large--scale magnetic field so that
a similar suppression of energy transfer is expected to occur.
In other words, fluctuating magnetic fields 
will inhibit the energy transfer from large--scale to small--scale
magnetic fields (responsible for turbulent diffusion), even when 
the latter is far 
below equipartition value. This link between small and large  scale 
magnetic field dynamics is indeed the very feature that is missing 
in LV, where a global reconnection rate is considered to be a simple sum of
local reconnection events, without depending on either $\langle B \rangle$
or $R_m$. That is, even if one local reconnection event 
may proceeds fast, the energy transfer from large--scale to small--scale 
is suppressed inversely with $R_m$, preventing many local
reconnection events for a large $R_m$ and fixed large--scale field
strength. Thus, the global reconnection rate is very likely to be reduced for 
large $R_m$. 

The purpose of this paper is to determine the global reconnection rate
by treating the dynamics of large and small--scale magnetic fields 
in a consistent way. The key idea is to compute the effective
dissipation rate of a large--scale magnetic field (turbulent
diffusivity) by taking into account small--scale field backreaction 
and then to use 
Sweek-Parker type balance relations to obtain the global reconnection rate.  
Since magnetic fields across current sheets are not 
always strictly antiparallel in real systems, we assume that only one 
component of the magnetic field (e.g., poloidal or horizontal field) changes 
its sign across the current sheet (see Fig. 2). The other component 
(e.g., axial field) is assumed to be very strong compared to the 
poloidal component. A strong axial magnetic field avoids the 
null point problem inherent in SP slab model, justifying the 
assumption of incompressibility of the flow in the poloidal (horizontal) 
plane. Such a magnetic configuration is ideal for the application 
of so--called 3D reduced MHD (3D RMHD) (Strauss 1976). 
In 3D RMHD, the conservation of the mean square 
potential is linearly broken due to the propagation of Alfv\'en waves
along an axial field, but preserved by the nonlinearity.
As we shall show later, the latter effect introduces additional 
suppression in the effective dissipation of a large--scale 
magnetic field compared to 2D MHD. 
We also discuss the 2D MHD case which can be recovered from our results
simply by taking the limit $B_0 \to 0$, where $B_0$ is a axial magnetic
field.

To be able to obtain analytic results, we adopt the following two methods. 
The first is a quasi--linear closure together using $\tau$ approximation by 
assuming the same correlation time for fluctuating velocity and magnetic fields
employing unity magnetic Prandtl number.
The second is an eddy-damped fluid model, based on large viscosity (Kim 1999),
which may have relevance in Galaxy where $\nu \gg \eta$. 
In this model, the nonlinear 
backreaction can be incorporated consistently, without
having to invoking the presence of fully developed MHD turbulence,
or assumptions such as a quasi--linear closure or $\tau$ approximation.
In both models, the isotropy and homogeneity of
turbulence is assumed in the horizontal (poloidal) plane since the 
reduction 
in effective dissipation of a large--scale poloidal magnetic field is 
likely to occur when its strength is far below the equipartition value. 
The effect of hyper-resistivity is incorporated in our analysis. This 
can potentially accelerate the dissipation of a large--scale poloidal
magnetic field. 

The paper is organized in the following way.
In \S 2, we set up our problem in 3D RMHD and provide the quasi--linear
closure using $\tau$ approximation where the flux is
estimated in a stationary case. Section 3 contains a similar analysis 
for an eddy-damped fluid model.  The global reconnection rate for both
models is presented in \S 4. Our main conclusion and discussion
is found in \S 5. 

\section{QUASI--LINEAR MEAN FIELD EQUATIONS}

We assume that a strong constant axial magnetic field $B_0$ is aligned 
in the $z$ direction and that a poloidal (horizontal) magnetic field $\B_H$
lies in the horizontal $x$-$y$ plane, as shown in Fig. 2. The subscript $H$ 
denotes horizontal direction. The total magnetic field is then expressed 
as $\B=B_0 {\hat z} + \B_H = B_0 {\hat z} + \curl \psi {\hat z}$, in terms 
of a parallel component of the vector potential $\psi$ 
(i.e., $\B_H = \curl \psi {\hat z}$). According to the 
RMHD ordering, the flow in the horizontal plane $\bu$ 
is incompressible and therefore can be written using a scalar
potential $\phi$ as $\bu = \curl \phi {\hat z}$. 
Then, the equations governing 3D RMHD are (see Strauss 1976):
\beqar
\p_t \psi + \bu \cdot \nabla \psi &=& \eta \lap \psi + B_0 \p_z \psi\,,
\label{eq1}\\
\p_t \lap \phi + \bu \cdot \nabla \lap \phi &=& \nu \lap \lap \phi 
+ \B\cdot \nabla \lap \psi\,,
\label{eq2}
\eeqar
where $\eta$ and $\nu$ are Ohmic diffusivity and viscosity, respectively. 
For the quasi--linear closure, unity magnetic Prandtl number
($\eta=\nu$) will implicitly be assumed.
In comparison with 2D MHD, the equation for the vector potential
contains an additional term $B_0 \p_z \phi$, which reflects the propagation 
of Alfv\'en wave along the axial magnetic field $B_0 {\hat z}$. Due
to this additional term, the conservation of the mean square 
potential is broken in 3D RMHD, albeit only linearly. In other words,
the nonlinear term in equation (2) conserves $\laa \psi^2 \ra$
since $\langle \bu \cdot \nabla \psi^2 \rangle = \nabla \cdot \langle 
\bu \psi^2 \ra = 0$, assuming that boundary terms vanish (cf. Blackman
\& Field 2000). Similarly, the momentum
equation contains an additional term $B_0 \p_z \lap \psi$. These 
additional terms 
are proportional to the wavenumber $k_z$ along $B_0 {\hat z}$. Thus, the
2D case can be recovered by taking $k_z \to 0$ or $B_0 \to 0$. 
Note that due to a strong axial field $B_0{\hat z}$, the vertical wavenumber
$k_z$ is much smaller than horizontal wavenumber $\k_H=k_x {\hat x}
+ k_y {\hat z}$; specifically, the 3D RMHD ordering implies that 
$k_z/k_H \sim B_H/B_0 \sim \epsilon \ll 1$. 

We envisage a situation where large--scale magnetic fields
with a horizontal component $\B_H = \langle \B_H \rangle=\curl 
\laa \psi \ra {\hat z}$ are embedded in a turbulent background.
The turbulence can be generated by an external forcing, for instance.
The horizontal component of a large--scale magnetic field 
$\laa \B_H\ra $ flows to form a current sheet of 
thickness $\Delta$ in the horizontal plane, so $\laa \B_H\ra$ 
changes sign across the current sheet. As reconnection proceeds, 
small--scale flows as well as magnetic fields are generated within 
the current sheet. It is reasonable to model the physical processes 
within a current sheet as well as the background turbulence
by an (approximately) isotropic and homogeneous 
turbulence with fluctuating velocity $\bu$ and magnetic field 
$\bb = \curl \psi'{\hat z}$.
Here the assumption of isotropy is justified since 
$\laa B_H\ra^2 \ll \laa u^2 \ra$, i.e. the reconnecting field
is taken to be weak. 

Outside the reconnection region, there are large--scale inflow 
and outflow in addition to the background turbulence. Thus,
to obtain SP-like balance relations, small--scale flow as well
as large--scale flow should be incorporated. However, since small--scale 
velocity is assumed to be homogeneous and isotropic, there is no net
contribution from the fluctuating velocity to mass continuity.
Effectively, the small--scale velocity does not appear
in the momentum balance either. 
However, Ohm's law (magnetic energy balance) now contains an 
additional term due to the correlation between fluctuating fields 
$\laa \bu \times \bb \ra$,
leading to turbulent diffusitivy (effective dissipation
rate), which then effectively changes the Ohmic diffusivity to the sum
of Ohmic diffusivity and turbulent diffusivity inside current sheet.
Therefore, similar balance relations to the original SP 
hold in our case
as long as the Ohmic diffusivity is replaced by the total diffusivity. 

To recapitulate, homogeneous and isotropic turbulence is assumed
to be present with magnetic fields $\B_H = \langle \B_H \rangle + 
\bb$ ($\laa \bb \ra =0$) and small--scale velocity $\bu$ ($\laa \bu \ra
 = \laa \phi \ra=0$). 
Once the effective dissipation rate of $\laa \B_H \ra$ within the 
reconnection zone is computed, it will be used to determine the 
reconnection velocity $v_r$ through SP balance relations
by using the total diffusivity in place of Ohmic diffusivity.

\subsection{\it Mean Field Equation}

The evolution equation for $\psi$ is obtained by taking the average  
of the above equation as:
\beqar
\p_t \langle \psi \rangle + \langle \bu \cdot \nabla \psi' \rangle 
&=& \eta \lap \langle \psi \rangle\,.
\label{eq3}
\eeqar
Note that although equation (4) does not exhibit an explicit dependence
on $B_0$, it does depend on $B_0$ through the flux $\Gamma_i \equiv \langle 
u_i \psi'\rangle$. To compute the flux $\Gamma _i$, we first do a 
quasi--linear closure of $\langle \bu \cdot \nabla \psi' \rangle$.

The effect of the backreaction can be incorporated in the flux
$\Gamma_i$ by considering the change in flux $\Gamma_i$ to be due to the 
change in the velocity as well as the fluctuating magnetic field. That is, 
we can rewrite the flux as
\beqar
\Gamma_i &=& \epsilon_{ij3} \langle \p_j \phi \psi'\rangle =
\epsilon_{ij3} \langle \p_j \phi \delta \psi' 
-\delta \phi \p_j \psi'\rangle\,,
\label{eq4}
\eeqar 
where unity magnetic Prandtl number is assumed for the 
equal splitting between $\langle \p_j \phi \delta \psi' \rangle$
and $\langle \delta \phi \p_j \psi'\rangle$; the latter 
essentially takes the backreaction to be as important as the kinematic 
contribution.

\subsection{\it Fluctuations}

>From equations (2) and (3), we can write the equation for the 
fluctuations in the following form.
\beqar
(\p_t + \bu \cdot \nabla) \psi' -\laa \bu \cdot \nabla \psi'\ra
&=&
- \bu \cdot \nabla \laa \psi \ra+ \eta \lap \psi' + B_0 \p_z \psi'\,,
\nonumber \\
(\p_t +\bu \cdot \nabla) \lap \phi -\laa \bu \cdot \nabla \lap \phi \ra
&=& \nu \lap \lap \phi + B_0\p_z  \lap \psi'
+ \laa B_H\ra \cdot \nabla_H  \lap \psi'
+ \bb \cdot \nabla_H  \lap \laa \psi \ra\,.
\nonumber
\eeqar
Here we have assumed that there is no large--scale flow in the current
sheet.
To estimate $\delta \phi$ and $\delta \psi'$ in equation (5), we
introduce a correlation time $\tau$ that represents the overall
effect of inertial and advection terms on the left hand side
of the above equations. That is, we approximate 
$(\p_t + \bu \cdot \nabla) \psi' -\langle \bu \cdot \nabla \psi'\rangle 
\equiv \tau^{-1} \psi'$, and
$(\p_t +\bu \cdot \nabla) \lap \phi -\langle \bu \cdot \nabla \lap \phi
\ra \equiv \tau^{-1} \lap \phi$, where the same correlation time 
$\tau$ is assumed 
for both the fluctuating flow and magnetic field due to unity magnetic
Prandtl number. Then, $\delta \phi$ and $\delta \psi'$ in
equation (5) can be estimated from the above equations as follows: 
\beqar
\delta \psi' &=&\tau\left[B_0 \p_z \phi' - \epsilon_{ij3} \p_j \phi' \p_i
\langle \psi\rangle\right]\,,
\label{eq5}\\
\delta \lap \phi &=& \tau\left[ B_0 \p_z \lap \psi' + \epsilon_{ij3} \p_j 
\langle \psi \rangle \p_i \lap \psi' + \epsilon_{ij3}\p_j \psi'
\p_i \lap \langle \psi \rangle \right]\,.
\label{eq6}
\eeqar
In Fourier space, the above equations take the following form:
\beqar
\delta \psi'(\k)
&=& \tau \bigl[B_0 ik_z \phi(\k)
+ \epsilon_{ij3} \int d^3k' k'_j\phi(\k')(k-k')_i
\langle \psi(\k-\k')\rangle \bigr]\,,
\label{eq7}\\
\delta \phi(\k)
&=& i \tau \biggl[B_0 k_z \psi'(\k)
+ i \epsilon_{ij3} {1\over k^2} 
\int d^3k' \left[(k-k')_j k'_i k'^2 + k'_j(k-k')_j (\k-\k')^2
\right] \psi'(\k') 
\nonumber \\
&&\hskip6cm \times \langle \psi(\k-\k')\rangle \biggr]\,.
\label{eq8}
\eeqar
Note that in principle, the correlation time can be a function of 
the spatial scale,
or the wavenumber, i.e., $\tau = \tau_{{\k}}$. Nevertheless, for
the notational simplicity, we have taken $\tau$ to be
a constant by assuming that the variation  of $\tau_{\k}$ in ${\k}$ is 
small or that the small--scale fields possess a characteristic
scale with a small spread in $\k$. Our final result will not fundamentally
change when the scale dependence of $\tau$ is incorporated. 

The flux $\Gamma_i$ can readily be computed once the statistics of 
small--scale magnetic field and the velocity are specified. As mentioned 
earlier, the statistics of both fluctuations are assumed to be homogeneous and
isotropic in the $x$-$y$ plane. We further assume that the former
is homogeneous and reflectionally symmetric in the $z$ direction
with no cross correlation between horizontal and vertical components,
thereby eliminating a helicity term. The absence of helicity terms 
rules out a possibility of a mean field dynamo in our model. 
Note that due to the presence of a strong axial field $B_0 {\hat z}$, 
the correlation functions cannot be everywhere isotropic. 
Specifically, the correlation functions at equal time 
$t$ are taken to have the form:
\beqar
\langle \psi'(\k_1,t) \psi'(\k_2,t) \rangle
&=& \delta (\k_1-\k_2) {\overline \psi}(k_{1H}, k_{1z})\,,
\label{eq9}\\
\langle \phi(\k_1,t) \phi(\k_2,t) \rangle &=& \delta (\k_1-\k_2) 
{\overline \phi}(k_{1H},k_{1z})\,, \label{eq10}
\eeqar
where ${\overline \psi}(k_{1H},k_{1z})$ and ${\overline \phi}(k_{1H},k_{1z})$
are the power spectra of $\psi'$ and $\phi$, respectively. These depend
on only the magnitude of horizontal wavenumber $k_{1H}= \sqrt{k_{1x}^2
+ k_{1y}^2}$ and vertical wavenumber $k_{1z}$. Finally, we assume
that $\laa \phi \psi' \ra = 0$, which can be shown to be equivalent 
to excluding the generation of a large--scale flow by the Lorentz force.

Straightforward but tedious algebra using equations (8)--(11)
in equation (5) leads to the following expression for the flux
(the details are given in Appendix A):
\beqar
\Gamma_i
&=& - {\tau \over 2} \left[ (\langle u^2 \rangle
- \langle b^2 \rangle) \p_i \langle \psi \rangle
- \langle \psi'^2 \rangle \p_i \lap \langle \psi \rangle \right]\,,
\label{eq11}
\eeqar
where $\langle u^2 \rangle =\int d^3 k k^2 {\overline \phi}(\k)$,
$\langle \psi'^2 \rangle =\int d^3k {\overline \psi}(\k)$, and
$\langle b^2 \rangle =\int d^3k k^2 {\overline \psi}(\k)$.
The first term on the right hand side of equation (12) represents the 
kinematic turbulent diffusion by fluid advection of the flux; 
the second represents the flux coalescence due to the backreaction 
of small--scale magnetic fields with the (negative) diffusion coefficient
proportional to the small--scale magnetic energy $\laa b^2 \ra$. 
The third term is the hyper-resistivity, reflecting the contribution
to $\Gamma_i$ due to 
the gradient of a large--scale current $\laa J \ra =-\lap \laa \psi \ra$.
($J {\hat z} = \curl {\B}_H$).
Note that the value of hyper-resistivity, being proportional to mean square 
potential, is related to the small--scale magnetic energy
as $\laa \psi'^2 \ra =  L_{bH}^2 \laa b^2 \ra$, where
$L_{bH}$ is the typical horizontal scale of $\bb$.
Thus, the negative magnetic diffusion (second) term and hyper-resistivity
(third) term are closely linked through the small--scale magnetic
energy $\laa b^2 \ra$. Indeed, the negative diffusivity and
hyper-resistivity together conserve total $\laa \psi'^2 \ra$, while
shuffling the $\laa \psi'^2\ra$ spectrum toward large scales.  
We now put equation (12) in the following form:
\beqar
\laa b^2 \ra
&=& {2\Gamma_i/\tau + \laa u^2 \ra \p_i \laa \psi \ra
\over \p_i \laa \psi \ra + L_{bH}^2 \p_i \lap \laa \psi \ra}\,,
\label{eq12}
\eeqar
where no summation over the index $i$ occurs.

\subsection{\it Stationary Case: $\p_t\langle \psi'^2 \rangle=0$}

To compute the flux $\Gamma_i$, we need an additional relation
between $\laa b^2\ra$ and $\Gamma_i$ besides equation (13). This can 
be attained by imposing a stationarity condition on $\laa \psi'^2 \ra$. 
The stationarity of
fluctuations is achieved in a situation where the energy transfer
from large--scale fields balances the dissipation of fluctuations
locally, as is usually the case in the presence of an external forcing
and dissipation. To obtain this relation, we multiply the equation 
for $\psi'$ by $\psi'$ and then take the average
\beqar
{1\over 2} \p_t \langle \psi'^2 \rangle 
+ \epsilon_{ij3}\langle \p_j \phi \psi'\rangle \p_i \langle \psi \rangle
&=& -\eta \langle (\p_i \psi')^2 \rangle + 
B_0 \langle \psi' \p_z \phi \rangle \,.
\label{eq13}
\eeqar
Here, the integration by parts was used
assuming that there are no boundary terms. We note that 
either when the stationarity
condition is not satisfied or when boundary terms do not
vanish, there will be a correction to our results
(Blackman \& Field 2000). 
When $\langle \psi'^2 \rangle$ is stationary, the first term
on the left hand side of equation (14) vanishes, simplifying 
the equation that relates $\langle b^2 \rangle$ to
$\Gamma_i = \laa u_i \psi' \ra = \epsilon_{ij3}\laa \p_i\phi \psi'\ra$ to
the form:
\beqar
\langle (\p_i\psi')^2 \rangle &= &\langle b^2 \rangle
= {1\over \eta}\left[-\Gamma_i \p_i \langle \psi \rangle
+ B_0 \laa \psi' \p_z \phi\ra \right]\,.
\label{eq14}
\eeqar
Note that in 2D MHD ($B_0=0$), the flux is proportional to  $\eta 
\laa b^2 \ra$. This balance reflects the conservation of $\laa \psi^2 \ra$,
which is damped only by Ohmic diffusion. 
The second term on the right hand side of equation (15) can be
evaluated in a similar way as for $\Gamma_i$, i.e., 
by writing 
\beqar
\laa \psi' \p_z \phi\ra
&=& \laa \delta \psi' \p_z \phi - \p_z \psi' \delta \phi\ra\,,
\label{eq15}
\eeqar
and then by using equations (8)--(11). Omitting the intermediate
steps (see Appendix A for details), the final result is
\beqar
\laa \psi' \p_z \phi\ra
&=& \tau B_0 [\aalpha \laa u^2 \ra - \bbeta \laa b^2 \ra ]\,.
\label{eq16}
\eeqar
Here
\beqar
\aalpha &\equiv & 
\int d^3k k_z^2 {\overline \phi}(\k)/ \int d^3k k_H^2 {\overline \phi}(\k)\,,
\label{eq17}\\
\bbeta &\equiv &
\int d^3k k_z^2 {\overline \psi}(\k)/ \int d^3k k_H^2 {\overline \psi}(\k)\,,
\label{eq18}
\eeqar
and $k_H^2=k_x^2+k_y^2$. If the characteristic
horizontal and vertical scales of $\bu$ are $L_{vH}$ and $L_{vz}$, and if
those of $\bb$ are $L_{bH}$ and $L_{bz}$, then $\aalpha$ and $\bbeta$ can
be expressed in terms of these characteristic scales as:
\beqar
\aalpha = {L_{vH}^2\over L_{vz}^2}\,, &&
\bbeta = {L_{bH}^2\over L_{bz}^2}\,.
\label{eq19}
\eeqar
Insertion of equation (17) into (15) gives us
\beqar
\langle b^2 \rangle
&=& {1\over \eta}\left[-\Gamma_i \p_i \langle \psi \rangle
+ \tau \aalpha B_0^2 \laa u^2 \ra\right]/
\left(1+{\tau \bbeta\over \eta} B_0^2 \right) \,. 
\label{eq20}
\eeqar
Thus, from equations (13) and (21), we obtain 
\beqar
\Gamma_i
&=& -{\tau \over 2} \laa u^2 \ra 
{1 + {\tau \over \eta} B_0^2 (\bbeta-\aalpha)
+{\tau  L_{bH}^2 \over \eta} \aalpha B_0^2 
|{\p_i \lap \laa \psi \ra \over \p_i \laa \psi \ra}|
\over
1 + {\tau\over \eta}
\left[{1\over 2}  \laa B_H\ra^2 + \bbeta B_0^2
- {L_{bH}^2 \over 2}  \laa J\ra ^2 
\right]} \p_i\laa \psi \ra\,,
\label{eq21}
\eeqar
where ${ J}{\hat z} = \curl \B_H$ and the integration by part is 
used to express $\p_i \laa \psi \ra \p_i \lap \laa \psi \ra
= - (\lap \laa \psi \ra)^2 = -\laa J\ra ^2 < 0$.
Note the last term in the numerator and denominator 
in equation (22) comes from the hyper-resistivity. 
Equation (22) is the flux in 3D RMHD, which generalizes the 2D MHD 
result (Cattaneo \& Vainshtein 1991; Gruzinov \& Diamond 1994). 
Several aspects of this result are of interest.
First, in the limit 
as $\B_0 \to 0$ and $\laa \B_H \ra \to 0$ ($\laa J \ra \to 0$), the flux 
reduces to the kinematic value $\Gamma_i = -\eta_k \p_i \laa \psi \ra$,
with the kinematic turbulent diffusivity $\eta_k = \tau \laa u^2 \ra /2$. 
This corresponds to the 2D hydrodynamic result where the effect of
the Lorentz force is neglected. The full 2D MHD result can
be obtained by taking the limit $\B_0 \to 0$ in equation (22),
which will reproduce equation (1). This agrees  
with the well--known result on the suppression of flux 
diffusion in 2D (Cattaneo \& Vainshtein 1991; Gruzinov \& Diamond 1994).

Another interesting case may be the limit $\laa \B_H \ra \to 0$.
In fact, this limit can be shown to be consistent with the ordering 
of 3D RMHD as follows. 
First, note that 3D RMHD ordering 
($k_z/k_H \sim B_H/B_0 \sim \epsilon < 1$) requires 
$\bbeta B_0^2 \sim \laa B_H^2 \ra$. 
Since $\laa B_H \ra^2 \ll \laa B_H^2 \ra \sim \laa b^2 \ra$, 
we expect that $\bbeta B_0^2 \sim \laa b^2 \ra \gg \laa B_H \ra^2$. 
Furthermore, $L_{bH}^2 \laa J \ra^2 \sim (L_{bH}/L_{BH})^2 \laa B_H \ra ^2
< \laa B_H \ra^2$, where $L_{BH}$ is the characteristic scale of
$\laa B_H \ra$. Thus, the dominant term in the square brackets in
the denominator of equation (22) is $ \bbeta B_0^2 \sim \laa b^2 \ra$.
That is, the effect of $B_0$ seems to be stronger
than that of $\laa \B_H \ra$ in 3D RMHD. 

Finally, to determine whether $\B_0$ enhances the flux or not, 
we note that $\aalpha -\bbeta $ in equation (22) can be taken to be zero,
since the scales for $\bb$ and $\bu$ are likely to be comparable 
in this model, which employs unity magnetic Prandtl number. 
Then, we estimate the last term in the numerator, due to 
hyper--resistivity, to be 
$\tau \laa b^2 \ra L_{bH}^2/(\eta L_{BH}^2) \sim
(L_{bH}/L_{BH})^2 R_m$ where 
$\bbeta B_0^2 \sim \laa b^2 \ra$ and
$\laa b^2 \ra \sim \laa u^2 \ra$ are used. 
If $(L_{bH}/L_{BH})^2 \sim R_m^{-1}$, this term will be of
order unity. 
Note $R_m = ul/\eta$ is the magnetic Reynolds number, with $u$ and
$l$ being the characteristic amplitude and length scale of the
velocity. 
Therefore, equation (22) indicates that the flux is reduced on account of
the strong axial magnetic field $B_0$ as well as the horizontal 
reconnecting field $\laa \B_H \ra$. 
The above analyses will be used 
in \S 4.1 in order to estimate the effective dissipation
and global reconnection rate.

\section{EDDY-DAMPED FLUID MODEL}
The analysis performed in the previous section introduced an arbitrary 
correlation time $\tau$ that is assumed to be the same for both small--scale 
velocity and small--scale magnetic fields. Moreover, the quasi--linear 
closure is valid 
strictly only when the small--scale fields remain weaker than the
large--scale 
fields. In order to compensate for these shortcomings, we now consider 
an eddy-damped fluid model which is based a large viscosity
(Kim 1999). In this model, the fluid motion is
self--consistently generated by a forcing with a prescribed statistics
as well as by the Lorentz force, without having to assume the presence
of fully developed MHD turbulence, to invoke a quasi--linear closure,
or to introduce an arbitrary correlation time for the fluctuating
fields. This is the simplest model within which the nonlinear
effect of the back--reaction can rigorously be treated. Even though 
this model has limited applicability to a system with a large viscosity,
it could be quite relevant to small scale fields in Galaxy where
$\nu \gg \eta$. 
As shall be shown later, this model gives rise to an effective correlation 
time for the fluctuating magnetic fields that is given by the viscous
time $\tau_{\nu} = l_{bH}^2/\nu$, where $l_{bH}$ is the typical scale of
the magnetic fluctuations in the horizontal plane (cf eqs.\ [22] and [32]). 
Thus, in comparison with the $\tau$ approximation in the previous
section, this model is equivalent to replacing $\tau$ by $\tau_{\nu}$
despite the fact that some of detailed results for the two models are not the 
same.

{\subsection {\it Splitting of Velocity}} 
In a high viscosity limit with the fluid kinetic Reynolds number 
$Re =ul/\nu < 1$, the nonlinear advection term as well
as inertial term in the momentum equation can be neglected. Then, the 
linearity of the remaining terms in the 
momentum equation enables us to split the velocity into two components; the 
first --- random velocity --- is solely governed by the random forcing, and 
the second --- induced velocity --- is governed by the Lorentz force only. 
Specifically, we express the total velocity $\bu$ as $\bu = \bbv+\bbv'$,
where $\bbv$ and $\bbv'$ are the random and induced velocity, respectively,
and introduce velocity potential $\phi_0$ and $\phi_I$ as $\bbv = \curl 
\phi_0 {\hat z}$ and $\bbv' = \curl \phi_I {\hat z}$. Then, the
equations for these potentials are:
\beqar
0&=& \nu \lap \phi_0 +  F\,,
\label{eq22}\\
0&=& \nu \lap \phi_I + \B\cdot \nabla \lap \psi \,,
\label{eq23}
\eeqar
where the nonlinear advection term as well as the inertial term
is neglected since $Re<1$ is assumed.
In equation (23), $F$ is a prescribed forcing with known statistics.
Instead of solving equation (23) for $\phi_0$, we can
equivalently prescribe the statistics of the random velocity $\phi_0$ 
(or $\bbv$). Therefore, we assume that the statistics of random component
satisfies homogeneity and isotropy in the horizontal plane and 
homogeneity and reflectional symmetry in the
$z$ direction, respectively. Furthermore, we assume that it is delta
correlated in time. The correlation function is then given by:
\beqar
\langle \phi_0(\k_1,t_1) \phi_0(\k_2,t_2) \rangle 
&=& \delta (\k_1-\k_2) \delta(t_1-t_2){\overline \phi}_0
(k_{1H},k_{1z})\,,
\label{eq24} 
\eeqar
where ${\overline \phi}_0(k_{1H},k_{1z})$ is the power spectrum of $\phi_0$.
Note that $\tau_0 \laa \phi_0^2 \ra = \int d^3k {\overline \phi}(\k)$
and $\tau_0 \laa v^2 \ra = \int d^3kk^2 {\overline \phi}(\k)$,
where $\tau_0$ is the correlation time of $\bbv$ that is assumed to
be short.

On the other hand, the induced velocity can be constructed by solving 
equation (24)
for $\phi_I$ in terms of $\B$. This can easily be done in Fourier
space as:
\beqar
\phi_I(\k)
&=& {i\over \nu k^2 k_H^2} \left[B_0 k^2 k_H^2
+ i \epsilon_{ij3}\int d^3 k'(k-k')_j k_{Hi}' k_H'^2 \psi(\k-\k')
\psi(\k')\right]\,,
\label{eq25}
\eeqar
where $B_{Hi}(\k) = i \epsilon_{ij3} k_j \psi(\k)$ is used.
Note that the $\psi$ in the above equation contains both mean
and fluctuating parts. 

{\subsection {\it Magnetic Field}}

Both random and induced velocities are to be substituted in
equation (2) to solve for the magnetic field. Notice that
equation (2) then has a cubic nonlinearity, since the induced
velocity is quadratic in $\B$. We again assume that the magnetic field 
in the horizontal plane consists of mean and fluctuating components, 
i.e., $\psi = \laa \psi \ra + \psi'$ and that the fluctuation is 
homogeneous and isotropic in the $x$-$y$ plane and homogeneous and 
reflectionally symmetric in the $z$ direction, satisfying the same 
correlation function as equation (10).

To obtain equations for $\laa \psi\ra$ and $\laa \psi^2\ra$, we 
utilize the delta--correlation in time of $\bbv$ and
iterate equation (2) for small time intervals $\delta t$. 
Specifically, we use $\laa v_i(t_1) B(t)_j\ra=0$ for $t_1>t$
and $v \sim O((\delta t)^{-1/2})$ since 
$\laa v_i(t_1) v_j(t_2)\ra \propto \delta (t_1-t_2) \sim 1/\dt $, where
$\dt = t_1-t_2$. Then, for $\delta t \ll 1$, equation (2)
can be iterated up to order $O(\delta t)$ as:
\beqar
&&\psi(t+\dt) 
\nonumber \\
&&= \psi(t) + \dt \eta \lap \psi(t) + \int_t^{t+\dt}dt_1
\left[\epsilon_{ij3}\p_j\psi(t) \p_i \phi (t_1) + B_0 \p_z \psi(t_1)
\right] \nonumber \\
&&+{1\over 2} \epsilon_{ij3}\int_t^{t+\dt} dt_1 dt_2
\left[\epsilon_{lm3}\p_i \phi(t_1)\p_j[\p_m \psi(t) \p_l \phi(t_2)]
+ B_0 \p_i \phi(t_1) \p_{jz}\phi(t_2)\right] + O(\dt^{3/2})\,,
\label{eq26}
\eeqar
where $\psi$ and $\phi$ are to be evaluated at the same spatial
position $\x$. 

The mean field equation is obtained by substituting equation (26)
in (27), by taking the average with the help of equations (10)
and (25), and then by taking the limit $\dt \to 0$. 
The derivation is tedious and is outlined 
in Appendix B. Here, we give the final result
\beqar
\p_t \laa \psi \ra
&=& \eta \lap\laa \psi \ra + \left[{\tau_0\over 4} \laa v^2 \ra
- {1\over 2\nu} G\right] 
\lap \laa \psi \ra - {F \over \nu} \lap \lap \laa \psi \ra
\nonumber \\
&=& (\eta + \eta_M) \lap \laa \psi \ra - \mu \lap \lap \laa \psi \ra\,.
\label{eq27} 
\eeqar
Here $\tau_0$ is the short correlation time of random velocity $\bbv$
and  
\beqar
\eta_M &\equiv& {\tau_0\over 4} \laa v^2 \ra - {1\over 2 \nu} G 
\equiv \eta_k - {1\over 2 \nu} G \,,
\nonumber  \\
\mu &\equiv & {F\over \nu}\,,
\nonumber \\
G &\equiv & \int d^3k {k_H^2 \over k^2} {\overline \psi}(\k)
\simeq \laa \psi'^2 \ra \equiv \kappa \laa b^2 \ra\,,
\nonumber \\ 
F &\equiv & \int d^3k {k_H^2 k_z^2\over k^6} {\overline \psi}(\k)
\simeq {L_{bH}^4\over L_{bz}^2} G \equiv \gamma G\,,
\nonumber
\eeqar
where $\eta_k = \tau_0 \laa v^2 \ra/4$ is the kinematic diffusivity; 
$\kappa \equiv L_{bH}^2$ and $\gamma \equiv L_{bH}^4/L_{bz} ^2
= \kappa \bbeta$.
The above equation implies that the flux $\Gamma _i = 
\laa u_i \psi' \ra$ is given by
\beqar
\Gamma_i &=& -\eta_M \p_i \laa \psi \ra + \mu \p_i \lap 
\laa \psi \ra \,.
\label{eq28}
\eeqar  
Again, the two terms in $\eta_M$ are due to the kinematic 
turbulent diffusivity and backreaction. Note that the
kinematic diffusivity $\eta_k = \tau_0 \laa v^2 \ra/4$ now comes 
only from the random velocity, with $\tau_0$ being its correlation
time that can be prescribed. The backreaction term is proportional
to $\laa \psi'^2 \ra$, not $\laa b^2 \ra$ (cf. eq.\ [11]) and
inversely proportional to the viscosity $\nu$. It is because the cutoff
scale of the magnetic field $l_\eta$ is smaller than that of the 
velocity $l_\nu$ in this model so that for a larger $\nu$, there are 
magnetic modes over a larger interval of scale $l$ between $l_\eta$ 
and $l_\nu$ (i.e. $l_\eta<l<l_\nu$) where the velocity is absent 
due to viscous damping.  That is, the induced velocity (Lorentz force) 
cannot be generated on this scale ($l_\eta < l <l_\nu$) due to viscous 
damping, thereby weakening the overall effect of backreaction 
(see eq. [48]). 
Now, the last term in equation (29) is the contribution from
the hyper-resistivity $\mu$. It is interesting to see that $\mu$ 
is inversely proportional to $L_{bz}^2$ and thus 
vanishes as $L_{bz} \to \infty$ (or $\gamma \to 0$) which corresponds
to the 2D limit. Therefore, 
in this eddy-damped fluid model, the hyper-resistivity
term vanishes in two dimensions. It should be contrasted to the case
considered in the previous section where the hyper-resistivity,
being proportional $\laa \psi'^2 \ra$, survives in 2D MHD limit 
(see eq. [12]). 

For use later, we solve equation (29) for $\laa b^2 \ra$ yielding 
\beqar
\laa b^2 \ra
&=& {\Gamma_i + \eta_k \p_i \laa \psi \ra
\over {\kappa \over 2\nu} \p_i \laa \psi \ra + 
{\kappa \gamma \over \nu} \p_i \lap\laa \psi \ra}\,,
\label{eq29}
\eeqar
where again the summation over the index $i$ is not implied. 

\subsection{\it Stationary Case: $\p_t\laa \psi'^2 \ra=0$}

The additional relation between the flux $\Gamma_i$ and 
magnetic energy $\laa b^2\ra$ is obtained for the case of
stationary $\langle \psi'^2 \rangle$. To derive an equation 
for $\laa \psi^2 \ra$, we 
multiply equation (27) by itself, take average, and then 
take the limit of $\delta t \to 0$. After considerable 
algebra (see Appendix B), we obtain the following equation 
\beqar
\p_t \laa\psi'^2\ra + \p_t \laa\psi\ra^2
-2 \eta \left[-\laa (\p_i \psi)^2 \ra
+ \laa \psi \ra \lap \laa \psi \ra \right]
&=& B_0^2 \left[\aalpha \laa v^2 \ra - {2\over \nu} {\overline G}\right]\,,
\label{eq30}
\eeqar
where
\beqar
{\overline G}&\equiv & \int d^3 k {k_z^2\over k^2} {\overline \psi}(\k)
\sim {L_{bz}^2\over L_{bH}^2} G= \bbeta G\,,
\nonumber 
\eeqar
In a stationary case, equations (28), (30), and (31) lead us to
the following expression for the flux:
\beqar
\Gamma_i &=& - {\tau_0\over 4} \laa v^2 \ra 
{1 + {\kappa\over \eta \nu} B_0^2 (\bbeta-\aalpha)
+ {2 \kappa \gamma \over \eta \nu} \aalpha B_0^2 
\left |\p_i \lap \laa \psi \ra\over \p_i \laa \psi \ra \right|
\over 1 + {\kappa \over \eta \nu} 
\left[\bbeta B_0^2 + {1\over 2} \laa B_H \ra ^2 - \gamma \laa J\ra^2
\right]} \p_i \laa \psi \ra\,,
\label{eq31}
\eeqar
where ${ J}{\hat z} = \curl \B_H$, and 
$\p_i \lap \laa \psi \ra \p_i \laa \psi \ra = -(\lap\laa \psi \ra)^2
= -\laa J \ra^2 < 0$ is used. When the characteristic
scales of fluctuating velocity and magnetic field are 
comparable, or when only the ratios of vertical to horizontal scales
of the fluctuating velocity and magnetic fields are comparable, 
$\aalpha$ can be taken to be equal 
to $\bbeta$, simplifying the above expression.

It is worth considering a few interesting limits of equation (32).
First, in the limit $B_0 \to 0$ and $B_H \to 0$, equation (32)
again recovers the 2D hydrodynamic result with the kinematic
diffusivity $\eta_k = \tau_0 \laa v^2 \ra/4$. The limit
$B_0 \to 0$ leads to 2D MHD case where the suppression of
the turbulent diffusion arises from $\laa \B_H \ra$. 
In 3D RMHD, the dominant suppression in the flux comes from
$B_0$ when $\aalpha = \bbeta$, as discussed in \S 2.3. 

We note that the last term in the numerator and denominator
is due to the hyper-resistivity, 
which comes with a multiplicative factor $\gamma = L_{bH}^2
/ L_{bz}^2 \ll 1$. Therefore, the effect of hyper-resistivity
can be neglected as compared to other terms in equation (32).
Since $\gamma \to 0$ in 2D MHD, there is no contribution
from the hyper-resistivity to the flux in 2D in this model. 
The estimate of the effective
dissipation in this model is provided in \S 4.2. 

It is very interesting to compare equation (32) with (22). We recall
that in order to derive equation (22), the same correlation time $\tau$ 
was assumed for both fluctuating magnetic field and velocity, which 
appears in front of the mean magnetic fields $B_0$ and $\laa \psi \ra$
in equation (22). In contrast, $\tau_0$ in equation (32)
is the correlation time of the random component of the velocity,
which can be arbitrarily prescribed. Moreover, $\tau$ in front of
mean magnetic fields in equation (22) is now replaced by viscous
time scale $\tau_{\nu} = \kappa/\nu = L_{bH}^2/\nu$ in equation (32). 
The latter
represents the viscous time scale across the typical horizontal scale 
of fluctuating magnetic fields. Thus, as noted at the beginning
of this section, this viscous time $\tau_{\nu}$ replaces $\tau$ in 
the quasi--linear closure, which was assumed to be a parameter.

\section{RECONNECTION RATE}
In previous sections, the flux $\Gamma_i$ was derived by using a 
quasi--linear closure and an eddy-damped fluid model. Assuming the 
flux $\Gamma_i$ has a form proportional to $\p_i \laa \psi \ra$ in both 
cases (see eqs. [22] and [32]), it can be expressed in terms of 
the effective dissipation rate (or, turbulent 
diffusivity) $\eta_{eff}$ as follows:  
\beqar
\Gamma_i&=& -\eta_{eff} \p_i \laa \psi \ra\,.
\label{eq32}
\eeqar
Upon using equation (33), the mean field equation (4) then becomes 
\beqar
\p_t\laa \psi \ra &=& (\eta + \eta_{eff}) \lap \laa \psi \ra
\equiv \eta_T \lap \laa \psi\ra\,.
\label{eq33}
\eeqar
where $\eta_T\equiv \eta + \eta_{eff}$ is the total dissipation rate of  
the mean field.  The effective dissipation rate is the quantity that 
represents the overall decay rate of a large--scale magnetic field
due to both small--scale motions and magnetic fluctuations. 
That is, the dynamical system consisting of both small and
large scale fields can be represented by the evolution
of a large--scale field only when the effect of small--scale
fields is absorbed in this turbulent coefficient. 

In order to determine a global reconnection rate, we now invoke
the original SP type balance equations and use the total dissipation rate
in place of the Ohmic diffusivity (see \S 2): 
\beqar
v_r&=&{v_A \over \sqrt{v_A L / \eta_T}}\,.
\label{eq34}
\eeqar
Note that we have neglected a multiplicative correction factor to 
the reconnection rate in the eddy-damped model since its
dependence on $\nu$ is weak with $1/4$ power 
(for instance, see, Biskamp 1993). 
In the following subsections, we assume $\aalpha = \bbeta$ for
simplicity and estimate the reconnection rate via 
equation (35). Then, we briefly comment on the implication 
for reconnection assuming `Alfv\'enic turbulence', as Lazarian
and Vishniac (1999) did.

\subsection{{\it Using the Quasi-linear Result}}

The effective dissipation rate follows from equations (22) and
(32):  
\beqar
\eta_{eff}
&\simeq& {\tau \over 2} \laa u^2 \ra
{1 + 
{\tau  L_{bH}^2 \over \eta} \aalpha B_0^2 
|{\p_i \lap \laa \psi \ra \over \p_i \laa \psi \ra}|
\over
1 + {\tau\over \eta}
\left[{1\over 2} \laa B_H \ra^2 + \bbeta B_0^2
- {L_{bH}^2 \over 2}  \laa J\ra ^2 
\right]} \,,
\label{eq35}
\eeqar
after using $\aalpha = \bbeta$.
As shown in \S 2.3, the dominant term in the square brackets in
the denominator of equation (36) is $ \bbeta B_0^2 \sim \laa b^2 \ra$,
and the second term in the numerator is of order unity
if $L_{BH}^2 / L_{bH}^2 \sim R_m$. In that case,
$\eta_{eff}$ is roughly given by 
\beqar
\eta_{eff}&\sim& \eta_k {1\over 1+\tau \laa b^2 \ra / \eta}
\sim \eta_k {1\over 1+ 2 R_m \laa b^2 \ra / \laa u^2 \ra} \,,
\label{eq36}
\eeqar
where $\eta_k = \tau \laa u^2 \ra / 2$ is the kinematic value
of turbulent diffusivity in 2D and $R_m = \eta_k/\eta$.
In contrast to the 2D MHD result 
(eq. [1])), the equation (37) reveals that 
the effective diffusivity in 3D RMHD is more severely reduced
as $\laa b^2 \ra \gg \laa B_H \ra^2$ ($ = \laa B \ra^2$).
To determine the leading order contribution in equation (37),
we need to estimate $\laa b^2 \ra$. To do so, we
substitute equations (33) and (37) in (13) and use $L_{bH} < L_{BH}$
to obtain:
\beqar
\laa b^2 \ra &\sim & \laa u^2 \ra - {\eta \over \tau}
\sim \laa u^2 \ra \left[1-{1\over 2 R_m}\right]\,,
\label{eq37}
\eeqar
where $R_m = \eta_k /\eta = \tau \laa u^2 \ra / 2 \eta$
is used.
We note that $\laa b^2 \ra >0$ is guaranteed since
$\laa b^2 \ra > \laa B_H \ra^2$ (implying $R_m>1$) was assumed to 
derive the above equation.
Thus,
\beqar
{\tau \laa b^2 \ra \over \eta} 
&\sim & 2 R_m-{1}\,.
\nonumber
\eeqar
That is, for $R_m \gg 1$, $\tau \laa b^2 \ra /\eta \gg 1$. Insertion
of the above equation in (37) then gives us
\beqar
\eta_{eff} &\sim & \eta_k {1\over 2R_m} 
\sim {\eta\over 2} \,.
\label{eq38}
\eeqar  
In other words, to leading order, the effective dissipation rate is 
just that given by Ohmic diffusivity! Therefore, by inserting equation (39)
into (35) with $\eta_T = \eta + \eta_{eff}$, 
the reconnection rate is found to have the original
SP scaling with $\eta$, i.e.
\beqar
v_r&\sim &{v_A \over \sqrt{v_A L / \eta}}\,.
\label{eq39}
\eeqar

It is interesting to contrast this result to the 2D case where
$B_0 =0$. In that case, the dominant term in equation (36) 
is $\laa B_H \ra^2$, with $\eta_{eff} \sim \eta_k \laa u^2 \ra
/ R_m \laa B_H \ra ^2 \sim \eta \laa u^2 \ra / \laa B_H \ra^2
 \sim \eta u^2/v_A^2>\eta$, where $u$ is the typical velocity. 
Therefore, in 2D, the global reconnection rate becomes
\beqar
v_r&\sim &{v_A \over \sqrt{v_A L / \eta}} {u\over v_A}\,,
\label{eq40}
\eeqar
which is larger than SP by a factor of magnetic Mach number
$M_A = u/v_A$. 
Note that the reduction in the effective dissipation
of a large--scale magnetic field is more severe
in 3D RMHD than in 2D MHD by a factor of $\laa u^2 \ra /
\laa B_H \ra^2\sim \laa u^2 \ra/v_A^2$.  

\subsection{\it{Using the Eddy-Damped Fluid Model Result}}  

For an eddy-damped fluid model, equation (32) yields: 
\beqar
\eta_{eff} &=&  {\tau_0\over 4} \laa v^2 \ra
{1 + 
 {2 \kappa \gamma \over \eta \nu} \aalpha B_0^2 
\left |\p_i \lap \laa \psi \ra\over \p_i \laa \psi \ra \right|
\over 1 + {\kappa \over \eta \nu} 
\left[\bbeta B_0^2 + {1\over 2} \laa B_H \ra ^2 - \gamma \laa J\ra^2
\right]}\,,
\label{eq41}
\eeqar
after assuming $\aalpha = \bbeta$. We recall that the contribution from 
the hyper-resistivity comes with a multiplicative factor $\gamma = L_{bH}^2 
/ L_{bz}^2 \ll 1$ (vanishing in the 2D MHD limit) and thus can be 
neglected as compared to other terms in equation (42). Then, a similar 
estimation as in \S 4.1 simplifies equation (42) to
\beqar 
\eta_{eff} & \sim & \eta_k {1\over 1 + {\kappa \over \nu \eta}\laa b^2 \ra}\,,
\label{eq42}
\eeqar
where $\eta_k = \tau_0 \laa v^2 \ra/4$ is the kinematic value of the turbulent
diffusivity in 2D and $\kappa = L_{bH}^2$. To obtain the leading order 
behavior of equation (43), we estimate $\laa b^2 \ra$ with the help of
equation (30) to be 
\beqar
\laa b^2 \ra
&\sim & {\eta \nu \over \kappa} (2R_m-1)\,,
\label{eq43}
\eeqar
where $R_m = \eta_k/\eta$. 
By inserting equation (44) in (43), we obtain
\beqar
\eta_{eff} &\sim & {\eta_k \over 2 R_m} \sim {\eta\over 2}\,.
\label{eq44}
\eeqar
Thus, the reconnection rate is again given by 
\beqar
v_r&\sim &{v_A \over \sqrt{v_A L / \eta}}\,,
\label{eq45}
\eeqar
i.e., SP scaling with $\eta$ persists!

It is interesting to estimate $\laa b^2 \ra$ in equation (44)
by using  
\beqar
{\eta \nu \over  \kappa} &=&
\laa v^2 \ra {\eta  \over  \sqrt{\laa v^2 \ra} L_{bH}}
{\nu \over  \sqrt{\laa v^2 \ra} L_{bH}} \sim
 \laa v^2\ra {1\over R_m R_e}\,,
\label{eq46}
\eeqar
where $R_e = \sqrt{\laa v^2 \ra} L_{bH}/\nu$ is the fluid Reynolds number. 
Thus, equation (44) becomes 
\beqar
\laa b^2 \ra
&\sim & \laa v^2 \ra {1\over R_e} \left ( 2-{1\over R_m}\right) \,.
\label{eq47}
\eeqar
The above equation clearly demonstrates that $\laa b^2 \ra > \laa v^2 \ra$
for our model ($R_e < 1$) when $R_m > 1$, as pointed out near the
end of \S 3.2.

Finally, we note that in 2D limit with $B_ 0 \to 0$, 
the dominant term in the square brackets in the denominator of
equation (42) is $\laa B_H \ra^2$. 
Thus, $\eta_{eff} \sim \eta_k \laa v^2 \ra
/ R_e R_m \laa B_H \ra ^2 \sim \eta \laa v^2 \ra / R_e \laa B_H \ra^2
\sim \eta  u^2/R_e v_A^2>\eta$, where $u$ is the typical velocity.  
Therefore, in 2D, the global reconnection rate becomes
\beqar
v_r&\sim &{1 \over \sqrt{R_e}}
 {v_A \over \sqrt{v_A L / \eta}} {u\over v_A}\,,
\label{eq48}
\eeqar
where $u/v_A= M_A$ is the magnetic Mach number.
In comparison with equation (41), the global reconnection rate 
in this model is thus larger in the 2D limit (recall $R_e < 1$). 

{\subsection{\it{Alfv\'enic Turbulence}}
In Alfv\'enic turbulence (Goldreich \& Sridhar 1994; 1995; 1997), 
the equipartition between $\laa b^2 \ra$ 
and $\laa u^2 \ra$ is assumed from the start. It is to be contrasted to 
the present analysis in which the relation between $\laa b^2 \ra$ and 
$\laa u^2 \ra$ i.e., equations (38) and (49), follows from the condition
of stationarity of $\laa \psi'^2 \ra$ in the presence of $\B_0$
and $\laa \B_H\ra$.  As can be seen from equation (38),
in the quasi--linear closure with unity magnetic Prandtl number, 
exact equipartition is possible only for $\eta = 0$. In the 
eddy-damped fluid model, exact equipartition can never be satisfied
since the assumption $R_e<1$ implies $\laa b^2 \ra> \laa v^2 \ra$ 
when $R_m>1$ (see eq. [48])! 
Therefore, in general, stationarity of $\laa \psi'^2 \ra$ and exact 
Alfv\'enic equipartition cannot be simultaneously achieved. In other
words, if Alfv\'enic turbulence is assumed, $\laa \psi'^2 \ra$
cannot be stationary; if $\laa \psi'^2 \ra$ is stationary,
the turbulence cannot be in a state of Alfv\'enic equipartition. 

We easily confirm this in 2D MHD by quasi--linear closure.  The exact
equipartition ($\laa u^2 -b^2 \ra=0$) implies that the flux $\Gamma_i$ 
in equation (12) is given by hyper-resistivity only:
$\Gamma _ i = -\tau \laa \psi'^2 \ra \p_i \lap \laa \psi \ra/2$. 
Then, if we were to impose the stationarity of $\laa \psi'^2 \ra$,
equation (15) would indicate $\laa \psi'^2 \ra \tau \p \laa J\ra \laa B_H\ra 
= \eta \laa b^2 \ra$. Thus, 
\beqar
{\laa B_H \ra^2\over \laa b^2 \ra} R_m &\sim& 
\left({l_B\over l_b}\right)^2\,,
\label{eq49}
\eeqar
where $l_B$ and $l_b$ are the characteristic scales of $\laa \B_H\ra$
and $\bb$, respectively. Since $\laa B_H \ra^2 / \laa u^2 \ra \sim 1/R_m$
(with $\laa b^2 \ra \sim \laa u^2 \ra$)
and $(l_B/l_b)^2 \sim 1/R_m$ in 2D MHD, the relation (49) 
(for stationarity) cannot be satisfied. 

\section{CONCLUSION AND DISCUSSIONS}

In view of the ubiquity of turbulence in space and astrophysical plasmas, 
magnetic reconnection will likely occur in an environments with 
turbulence.  On the other hand, the reconnection itself generates 
small--scale fluctuation, feeding back the turbulence. Thus, it is
important to treat these two processes consistently, accounting for
the back reaction. Although LV argued that the local reconnection 
rate can be fast, they basically neglected the dynamic coupling between 
small and large scale fields, therefore leaving the issue of the global 
reconnection rate unresolved. The coupling between global and local
reconnection rates should be treated self consistently.
The aim of the present work was to shed some light on this issue 
by taking the simplest approach that is analytically tractable.

Our main strategy was to self--consistently compute the effective 
dissipation rate of a large--scale magnetic field within the current sheet 
by using stationarity of $\laa \psi'^2 \ra$ and then
use the effective dissipation rate in SP type balance relations to 
obtain the global reconnection rate. 
To avoid the null point problem associated with a 2D slab model,
we considered 3D RMHD, within which we can solidly justify the 
incompressibility
of the fluid in the horizontal plane. To facilitate analysis, two
models (methods) were employed, one being a quasi--linear closure 
with $\tau$ approximation and the other eddy-damped fluid model. 

The effective dissipation rate $\eta_{eff}$ that we obtained
generalizes the 2D MHD result (Cattaneo \& Vainshtein 1991;
Gruzinov \& Diamond 1994).
The quasi--linear closure predicted $\eta_{eff} \sim
 \eta_k/( 1+ 2 R_m \laa b^2 \ra / \laa u^2 \ra) \sim \eta/2$ 
(see eqs. [37]--[39]).
A similar result was obtained in the eddy-damped fluid model 
with $\eta_{eff} \sim \eta_k/( 1+ R_m R_e \laa b^2 \ra / \laa u^2 \ra) 
\sim \eta/2$ (see eqs. [43]--[45] and [47]).

The 2D result can simply be recovered from our results on the flux
by taking the limit $B_0 \to 0$. In that limit, 
$\eta_{eff} \sim \eta_k/( 1+ R_m \laa B_H \ra^2 / \laa u^2 \ra)$ 
according to the quasi--linear closure, 
consistent with previous work.
In the eddy-damped fluid model, 
$\eta_{eff} \sim \eta_k/( 1+ R_m R_e \laa B_H \ra^2 / \laa u^2 \ra)$.  

Since the effective dissipation rate $\eta_{eff}$ was found
to be the same in both models (in 3D RMHD), the global reconnection, 
obtained by invoking SP balance relations, was also the same
with the value $v_r\sim v_A/\sqrt{v_A L/\eta}$ in both models. This
result indicates that
the global reconnection rate is suppressed for large $R_m$ 
as an inverse power of $R_m^{1/2}$ such that the original SP
scaling with $\eta$ persists. Again, this persistent $\eta$ scaling 
results from the reduction in the effective
dissipation rate of a large--scale magnetic field for large $R_m$
mainly due to a strong axial magnetic field, with the effective
dissipation rate $\eta_{eff} \sim \eta$.  

Furthermore, in the 2D limit, the quasi--linear closure 
yielded the global reconnection rate $v_r \sim (v_A/\sqrt{v_A L/\eta})
(u/v_A)$, which is enhanced over SP by a factor 
of $M_A = u/v_A$ (note that $M_A$ can be large).
In contrast, the eddy-damped fluid model gave 
$v_r \sim \sqrt{R_e}^{-1} (v_A/\sqrt{v_A L/\eta}) (u/v_A)$.

The implication of these results for the LV scenario is that
no matter how fast local reconnection
events proceed, there is not enough energy transfer from large--scale
to small--scale magnetic fields to allow fast global reconnection. Therefore,
global reconnection cannot be given by a simple sum of the 
local reconnection events as LV suggested. We emphasize again
that the $\laa \psi'^2\ra$ balance played the crucial role
in determining the global reconnection rate consistently. 
Alternatively, an accurate calculation of the global reconnection
rates requires that (global) topological conservation laws
be enforced.  

The reduction in the effective dissipation in 2D is closely linked to
the conservation of mean square magnetic potential. In 3D RMHD,
the mean square of parallel component of potential is no longer an
ideal invariant due to the propagation of Alfv\'en waves along a
strong axial magnetic field. Nevertheless, the conservation
of mean magnetic potential is broken only linearly, which 
turned out to introduce additional suppression factors, as
compared to 2D. The 
interesting question is then how relevant these results would be in 
3D. The mean square potential is not an invariant of 3D MHD.
However, its conservation is broken nonlinearly, unlike 3D RMHD. Therefore,
the effective dissipation in 3D MHD may be very 
different from that in 3D RMHD, with the possibility that 
the former may not be reduced, at least, 
in the weak magnetic field limit (Gruzinov \& Diamond 1994; Kim 1999). 
Moreover, in 3D, there is a possibility of a dynamo, 
which brings in an additional transport coefficient (the $\alpha$ effect) 
into the problem. Some insights into the problem
of effective dissipation of a large--scale field in the presence
of a dynamo process might be obtained by considering a simple extension 
of the present 3D RMHD model by allowing a large--scale 
dynamo in the horizontal plane. Recall that this possibility was
ruled out in the present paper by assuming isotropy in the horizontal 
plane and reflectional symmetry in the axial direction, with no helicity
term (i.e., no correlation between horizontal and vertical component 
of fluctuations).

Considering some of limitations of the two models that were analyzed 
in the paper, such as the $\tau$ approximation, quasi--linear closure, 
low kinetic Reynolds number limit, etc, 
it will be very interesting to investigate our predictions via 
numerical computation. The stationarity of $\laa \psi'^2 \ra$
can be maintained as long as there is an energy source in the
system, such as an external forcing. By incorporating the proper
ordering required for 3D RMHD, one can measure the decay rate of
$\laa \B_H \ra$ to check our predictions for $\eta_{eff} \sim \eta$
(see eqs. [40] and [46]). 
Ultimately, a numerical simulation with a simple reconnection
configuration should be performed to measure a global reconnection rate 
as a function of $R_m$ as well as $B_0$ and $\laa B_H\ra$.
It will also be interesting to investigate non--stationary
states such as plasmoid formation (Forbes \& Priest 1983;
Priest 1984; Matthaeus \& Lamkin 1986).

\vskip1cm

We thank E. Zweibel for bringing this problem to our attention
and for many interesting discussions. We also thank E.T. Vishniac 
and A.S. Ware for stimulating conversations.
This research was supported by U.S. DOE FG03-88ER 53275. 
P.H. Diamond also acknowledges partial support from the National
Science Foundation under Grant No. PHY99-07949 to the Institute
for Theoretical Physics at U.C.S.B., where part of this work
was performed. E. Kim acknowledges partial support from
HAO/NCAR where part of this work was completed. 

\section*{Appendix A}
\renewcommand{\theequation}{A.\arabic{equation}}
\setcounter{equation}{0}

In this appendix, we provide some of steps leading to equations (12)
and (17). First, to derive equation (12), we 
let $\Gamma_i = \Gamma_i^{(1)} - \Gamma_i^{(2)}$, where
$\Gamma_i^{(1)}= \epsilon_{ij3} \langle \p_j \phi \psi'\rangle$ and
$\Gamma_i^{(2)}= \epsilon_{ij3} \langle  \phi \p_i \psi'\rangle$,
and begin with $\Gamma_i^{(1)}$.
\beqar
\Gamma_i^{(1)}&= &\epsilon_{ij3} \langle \p_j \phi \psi'\rangle
\nonumber \\
&=&\epsilon_{ij3}\int d^3 k_1 d^3 k_2 ik_{1j}\laa \phi(\k_1)
\delta \psi'(\k_2)\ra \exp{\{i(\k_1+\k_2)\cdot \x\}}\,.
\label{a1}
\eeqar
After inserting equation (8) in (A1) and using equation (11),
we can easily obtain
\beqar
\Gamma_i^{(1)}& 
=&-i \tau \epsilon_{ij3}\epsilon_{lm3}
\int d^3 k_1 d^3 k k_{1j} k_{im} k_l {\overline \phi}(\k_1)
\laa \psi(\k)\ra e^{i\k\cdot \x}
+ \tau \epsilon_{ij3} \int d^3 k k_{1j} k_{1z} 
B_0 {\overline \phi} (\k_1)
\nonumber \\
&=& -{\tau \over 2} \p_l\laa \psi \ra \delta_{il} \int d^3k_1 k_1^2
{\overline \phi}(\k_1)
= -{\tau \over 2} \laa u^2 \ra \p_i\laa \psi \ra \,.
\label{a2}
\eeqar
where $\laa u^2 \ra = \int d^3k_1 k_1^2 {\overline \phi}(\k_1)$.
To obtain the last line in equation (A2), we use the following
relations 
\beqar
\int d^3 k k_j k_m {\overline \phi}(\k)
&=& {1\over 2} \delta_{jm} \int d^3 k k^2 {\overline \phi}(\k)\,,
\nonumber \\
\int d^3 k k_j k_z {\overline \phi}(\k)
&=& 0\,, 
\label{a3}
\eeqar
which follows from the isotropy of $\phi$ in the $x$-$y$ plane,
and reflectional symmetry in the $z$ direction.
 
The second part, $\Gamma_i^{(2)}$, is calculated in a similar way.
\beqar
\Gamma_i^{(2)}&= &\epsilon_{ij3} \langle \phi \p_j  \psi'\rangle
\nonumber \\
&=&\epsilon_{ij3}\int d^3 k_1 d^3 k_2 ik_{2j}\laa \delta \phi(\k_1)
\psi'(\k_2) \ra \exp{\{i(\k_1+\k_2)\cdot \x\}}\,.
\label{a4}
\eeqar
We insert equation (9) in (A4) and use (10) to obtain
\beqar
\Gamma_i^{(2)}
&=& i\tau \epsilon_{ij3}
\biggl[-iB_0 \int d^3 k_1 k_{1z} k_{1j} {\overline \psi}(\k_1)
\nonumber \\
&&+\epsilon_{lm3}\int d^3k_2 d^3 k e^{i\k\cdot \x}
{1\over (\k+\k_2)^2}
\left[k_m k_{2l} k_2^2 + k_{2m} k_l k^2\right]
k_{2j} {\overline \psi}(\k_2) \laa \psi (\k) \ra\biggr]
\label{a5}
\eeqar
Since $\laa \psi \ra$ has a scale much larger than $\psi'$, 
$k_2 \gg k$ in the second integral on the right hand side. We thus expand 
the integrand of this second term and use the following isotropy relations:
\beqar
\int d^3 k k_j k_m {\overline \psi}(\k)
&=& {1\over 2} \delta_{jm} \int d^3 k k^2 {\overline \psi}(\k)\,,
\nonumber\\
\int d^3 k k_i k_j k_l k_m {\overline \psi}(\k)
&=& {1\over 8} \left(\delta_{ij} \delta_{lm} + \delta_{il} \delta_{jm}
+ \delta_{im} \delta_{jl}\right) \int d^3 k k^4 {\overline \psi}(k)\,,
\nonumber\\
\int d^3 k k_i k_z {\overline \psi}(\k)
&=& 0\,. 
\label{a6}
\eeqar
A bit of algebra then gives us
\beqar
\Gamma_i^{(2)}
&=& {\tau \over 2}
\left[-\laa b^2 \ra \p_i \laa \psi(\x)\ra
- \laa \psi'^2 \ra \p_i \lap\laa \psi (\x)\ra\right]\,.
\label{a7}
\eeqar
Thus, from equations (A3) and (A7), we obtain equation (12) 
in the main text.

Second, to derive equation (17), we again compute the correlation 
function on the right hand side of equation (16) in Fourier
space. The first term can be rewritten as:
\beqar
\langle \delta \psi' \p_z \phi \rangle
&=&\int d^3 k_1 d^3 k_2 ik_{1z}\laa \phi(\k_1)
\delta \psi'(\k_2)\ra \exp{\{i(\k_1+\k_2)\cdot \x\}}\,.
\label{a8}
\eeqar
Then, inserting equation (8) in (A8) and using equation (11) give
us 
\beqar
\langle \delta \psi' \p_z \phi \rangle
&=&\tau \biggl[
\int d^3 k_1 k_{1z} k_{1z} B_0 {\overline \phi}(\k_1) 
-\epsilon_{lm3} \int d^3 k_1 d^3 k k_{1z} k_{im} k_l {\overline \phi}(\k_1) 
\laa \psi(\k)\ra e^{i\k\cdot \x} \biggr]
\nonumber \\
&=& 
\tau B_0 \int d^3 k_1 k_{1z}^2  {\overline \phi}(\k_1)
= \tau B_0 \aalpha \laa u^2 \ra\,,
\label{a9}
\eeqar
where the isotropy and equation (18) were used to obtain the last line.  
Similarly, the second term on the right side of equation (16) is
easily calculated (in Fourier space) by using the isotropy condition.
The result is 
\beqar
\langle \p_z \psi' \delta \phi \rangle
&=& \tau B_0 \int d^3 k_1 k_{1z}^2  {\overline \psi}(\k_1)
= \tau B_0 \bbeta \laa b^2 \ra\,.
\label{a10}
\eeqar
Thus, equations (16), (A9), and (A10) yield equation (17), in the main
text.

\section*{Appendix B}
\renewcommand{\theequation}{B.\arabic{equation}}
\setcounter{equation}{0}

In this Appendix, we provide some of intermediate steps used to
obtain equations (28) and (31). 
For the mean field equation (28), we first take the average
of equation (27)
\beqar
\laa \psi(t+\dt) \ra 
- \laa \psi(t) \ra - \dt \eta \lap \laa \psi(t) \ra
&=& I_1 + I_2 + I_3
\label{b1}
\eeqar
where
\beqar
I_1&=& \int_t^{t+\dt}dt_1
\left[\epsilon_{ij3} \p_j\psi(t) \p_i \phi_I (t_1)\right]
\simeq \dt \epsilon_{ij3} \p_i \laa \p_j \psi(t) \phi_I(t)\ra
\equiv \dt \p_i \Delta_i \,,
\nonumber \\
I_2&=& \int_t^{t+\dt}dt_1 B_0 \p_z \laa \psi_I(t_1) \ra \simeq
\dt B_0 \p_z \laa \phi_I(t)\ra \,,
\nonumber \\
I_3 &=&
{1\over 2} \epsilon_{ij3}\int_t^{t+\dt} dt_1 dt_2
\laa \epsilon_{lm3}\p_i \phi_0(t_1)[\p_{jm} \psi(t) \p_l \phi_0(t_2)
+ \p_m \psi(t) \p_{jl} \phi(t_2)]
\nonumber \\
&&\hskip3cm + B_0 \p_i \phi_0(t_1) \p_{jz}\phi_0(t_2)\ra\,,
\label{b2}\
\eeqar
where $\Delta_i \equiv\epsilon_{ij3}\laa \p_j \psi(t) \phi_I(t)\ra$
and the smooth variation of the induced velocity $\phi_I$ in time was used
to approximate the time integrals in $I_1$ and $I_2$.  
To compute the averages, it is convenient to express the 
correlation function (24) in terms of $\bbv$ in real space as:
\beqar
\laa v_i(\x, t_1) v_j(\y, t_2) \ra
&=&  \delta(t_1-t_2) 
\left[T_L({\bf r}_H, r_z) \delta_{ij} + r_H {\p T_L \over \p r_H}
\left(\delta_{ij} - {r_{Hi} r_{Hj} \over r_H^2}\right)\right]\,,
\label{b3}
\eeqar
where ${\bf r} \equiv \y-\x$ and ${\bf r}_H$ is the horizontal component.
Note that the above relation implies
that at ${\bf r} = 0$, $\laa v_i (\x,t_1) v_j(\x,t_2)\ra = \delta (t_1-t_2) 
\delta_{ij} T_L(r=0)$ so that $T_L(0)=\tau_0 \laa v^2 \ra/2 = 2 \eta_k$.
Here $\tau_0$ is the short correlation time of $\bbv$ and $\eta_k
= \tau_0 \laa v^2 \ra/4$ is the kinematic diffusivity.
$\laa v_i(\x) v_j(\x)\ra$ is obviously related
to $\phi_0$ by $\laa \p_i \phi_0(\x,t_1)\p_l \phi_0(\y,t_2)\ra
= \delta_{il} \laa v_j(\x,t_1) v_j(\y,t_2)\ra
+  \laa v_i(\x,t_1) v_l(\y,t_2)\ra$.
By using $\bbv_j = -\epsilon_{ij3}\p_i \phi_0$ and $\laa \phi_0(t_1) 
\psi(t)\ra = 0$, $I_3$ is determined to be:
\beqar
I_3 &=& {1\over 2} \dt T_L(0) \lap \laa \psi \ra \,.
\label{b4}
\eeqar
$I_3$ represents the kinematic turbulent 
diffusivity. Next, to compute $I_2$, we take the inverse Fourier
transform of equation (26) and then take the average. Upon neglecting
$\p_z \laa \psi \ra\sim 0$, one can easily show that $I_2 = 0$.
Finally, $I_1$ contains the backreaction as well as hyper-resistivity.
To evaluate this term, we insert equation (26) in $\Delta_i$
to obtain
\beqar
\Delta_i
&=& \epsilon_{ij3} \laa \p_j \psi(t) \phi_I(t) \ra
\nonumber \\
&=& -{i\over \nu} \epsilon_{ij3} \epsilon_{lm3}
\int d^3 k_2 d^3 k' e^{i\k'\cdot\x}
{1\over (\k+\k')^2 (\k_H+\k_H')^2} {\overline \psi}(-\k)
P_{jlm} \laa \phi(\k')\ra\,,
\label{b5}
\eeqar
here 
\beqar
P_{jlm}
&\equiv & -k_j \left[k_m k_l' k_H'^2 + k_l k_m' k_H^2 \right]\,.
\nonumber
\eeqar
For notational convenience, we introduce $\q = \k_H$ so that
$q_3 = 0$.
Since the characteristic scale of $\laa \psi\ra $ is much larger 
than that of $\psi'$, $k'\ll k$ in equation (B5). Thus, we 
expand the integrand of equation (B5) to second order in $(k'/k)$ and
exploit the isotropy and homogeneity of $\psi'$ in the $x-y$ plane.
The latter implies equation (A5) (recall $\q = \k_H$) and  
also the following relations
\beqar
\int d^3k q_j q_l q_r k_n& = &\int d^3k q_j q_l q_r q_n\,,
\nonumber \\
\int d^3k q_j q_l k_z k_z &=& {1\over 2} \delta_{jl} \int d^3k q^2 k_z^2\,.
\label{b6}
\eeqar
Then, a fair amount of algebra reduces equation (B5) to
\beqar
\Delta_i
&=& -{1\over 2\nu}
\p_i \laa \psi \ra \int d^3 k {k_H^2 \over k^2} {\overline \psi} (\k) 
- {1\over \nu} \p_i \lap \laa \psi \ra 
\int d^3 k {k_H^2 k_z^2 \over k^6} {\overline \psi} (\k)
\nonumber \\
&=& -{G \over 2\nu}  \p_i \laa \psi \ra  
- {F\over \nu}  \p_i \lap \laa \psi \ra \,.
\label{b7}
\eeqar 
Note that there is no contribution from the first order term.
By inserting equation (B7) into (B1), by dividing both
sides by $\dt$, and then by taking the limit of $\dt \to 0$,
we obtain equation (28). 

Next, to derive equation (31), we multiply equation (27) by $\psi$
and then take average to obtain the following equation:
\beqar
\laa \psi^2(t+\dt)\ra - \laa \psi^2(t) \ra - 2 \eta \dt 
\laa \psi(t) \lap \psi(t) \ra
&=&
J_1 + J_2 + 2 J_3\,,
\label{b8}
\eeqar
where
\beqar
J_1 &\equiv & \int_t ^{t+\dt} dt_1 dt_2
\biggl\{\epsilon_{ij3} \epsilon_{lm3}
\laa \p_j \psi(t) \p_m(t) \p_i\phi_I(t_1)\p_l \phi_I(t_2)\ra
 + 2 B_0 \epsilon_{ij3} 
\laa \p_j \psi (t) \p_i \phi_I(t_1) \p_x \phi_I(t_2)\ra\biggr\}\,,
\nonumber \\
J_2 &=&\epsilon_{ij3}\int_t^{t+\dt} dt_1 dt_2
\laa \psi(t) \left[\p_i \phi_0(t_1) \epsilon_{lm3}
\p_j\left[\p_m\psi(t)\p_l\phi_0(t_2)\right]
+ B_0 \p_i\phi_0(t_1) \p_{jz}\phi_0(t_2)\right]\ra\,,
\nonumber \\
J_3&=&
\int_t^{t+\dt} dt_1
\laa \psi(t) \left[\epsilon_{ij3} \p_j \psi(t) \p_i \phi_I(t_1)
+ B_0 \p_z \phi_I(t_1)\right]\ra
\equiv \dt (J_{31} + J_{32})\,, 
\label{b9}
\eeqar
where $J_{31} \equiv \epsilon_{ij3} \laa \psi(t) \p_j \psi(t)
\p_i \phi_I(t)\ra$ and $J_{32} \equiv B_0 \laa \psi(t) \p_z \phi_I(t)\ra$. 

First, $J_1$ can easily be computed by using the correlation functions 
as
\beqar
J_1&=& \dt \left[T_L(0) \left[\laa b^2 \ra + \laa B_H\ra^2\right]
- B_0^2 \int d^3 k_z^2 {\overline \phi}(\k)\right]\,.
\label{b10}
\eeqar
Next, $J_2$ can be computed upon substituting
equation (26) and then splitting average by 
using $\laa \psi(t)\phi(t_1)\ra = 0$, with the result 
\beqar
J_2
&=&
\delta t T_L(0) \left[-\laa b^2 \ra + \laa \psi \ra \lap 
\laa \psi \ra \right]\,.
\label{b11}
\eeqar
For $J_3$, one can first show $J_{31} = 0$ due to isotropy. 
To compute $J_{32}$, we substitute equation (26) and use
$\laa \phi_I \ra =0$ to obtain 
\beqar
J_{32} &=&
-B_0 \int d^3 k_1 d^3 k \exp{\{i(\k_1+\k_2)\cdot \x\}}
{k_z \over \nu k_H^2 k^2}
\nonumber \\
&&\times \laa  \psi'(\k_1) \left[B_0 k_z k_H^2 \psi'(\k)
+i \epsilon_{ij3} \int d^3k' \psi(\k-\k')
(k-k')_j k'_i k_H'^2 \psi(\k')\right]\ra
\nonumber \\
&=& -{B_0\over \nu}
\biggl[B_0 \int d^3k_1 {k_{1z}^2\over k_1^2}{\overline \psi}(\k_1) 
\nonumber \\
&&+i \epsilon_{ij3} \int d^3 k d^3k_1
\exp{\{i(\k_1+\k_2)\cdot \x\}}{k_z\over k_H^2 k^2}
Q_{ij} {\overline \psi}(\k_1) \laa \psi(\k+\k_1)\ra\biggr]
\label{b12}
\eeqar
where $Q_{ij}\equiv -k_{1j}(k+k_1)_i (\k_H+\k_{1H})^2
- k_{1i}(k+k_1)_j k_{1H}^2$. By using the definition of ${\overline G}$
(see immediately after eq. [31]) and $\epsilon_{ij3} Q_{ij} 
= -\epsilon_{ij3}(k_H^2 + 2 \k_H\cdot \k_H)k_i k_{1j}$, equation (B12)
becomes
\beqar
J_{32}&=&
-{B_0\over \nu}
\left[ B_0 {\overline G}
-i \epsilon_{ij3} \int d^3 k' d^3 k e^{i\k'\cdot \x}
{k_z \over k_H^2 k^2 }
\left[k_H^2 + 2k_l(k'-k)_l\right]
k_i (k'-k)_j {\overline \psi}(-\k+\k')
\laa \psi(\k')\ra\right]
\nonumber \\
&=&-{B_0\over \nu}
\left[ B_0 {\overline G}
-  \epsilon_{ij3} \p_j d^3 k' \int d^3 k e^{i\k'\cdot \x}
{k_z k_i \over  k^2 }
\left[-1 + {2k_l k_l'\over k_H^2}\right]
{\overline \psi}(-\k+\k') \laa \psi(\k')\ra\right]\,.
\label{b13}
\eeqar
Now, since $k'\ll k$, we expand the integrand of equation (B13)
to second order in $k'/k$, in order to show that there is no
contribution from the second term in equation (B13) (to this order).
Therefore, $J_{32} = - B_0^2 {\overline G}/\nu$. 
Inserting $J_1$, $J_2$, and $J_3$ in equation (B8), dividing by
$\dt$, and then taking the limit $\dt\to 0$ finally yields 
equation (31).

\pagebreak

\clearpage

%\plotone{f1.ps}
\figcaption{
Sweet--Parker 2D slab configuration. 
$\Delta$ and $L$ are the thickness and length of the current sheet;
$\pm \B$ are reconnecting magnetic fields;
$v_r$ and $v_0$ are inflow (reconnection) and outflow velocities. 
\label{F1}}

%\plotone{f2.ps}
\figcaption{
Configuration in 3D RMHD. $\B_0$ is a strong axial magnetic
field pointing in the $z$ direction, and $\pm \B_H$ are reconnecting
(large--scale) magnetic fields in the $x$-$y$ plane.
Panel (a) shows the projection in the $x$-$y$ plane, and Panel (b) 
in the $y$-$z$ plane.
\label{F2}}

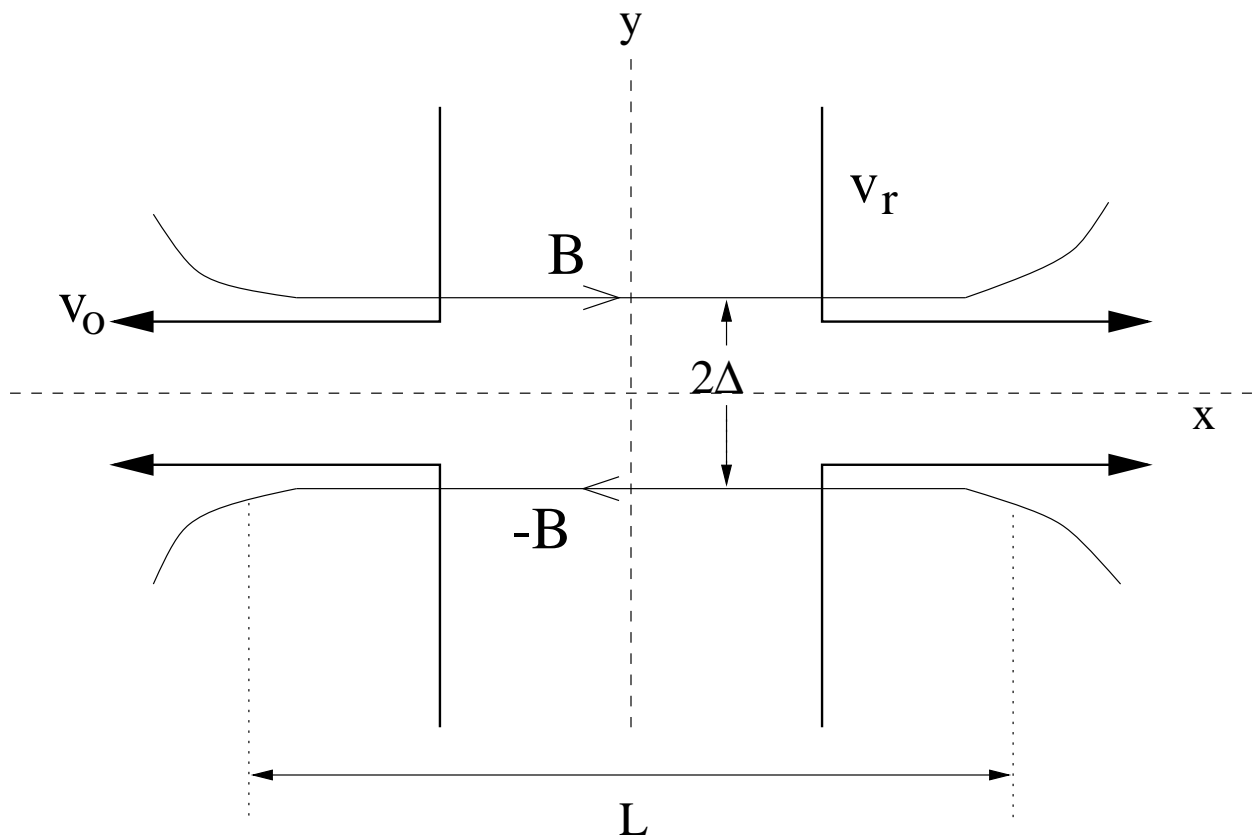
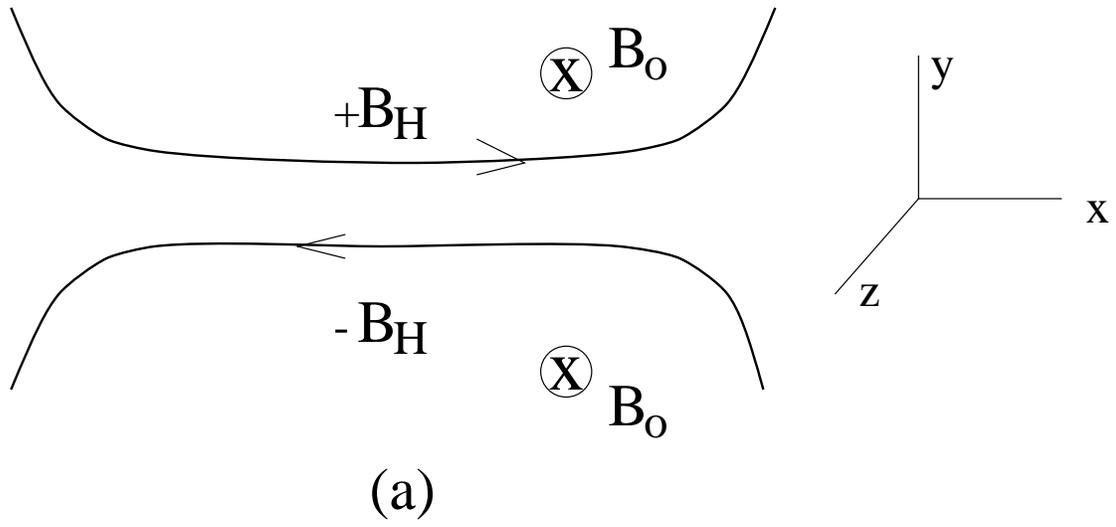
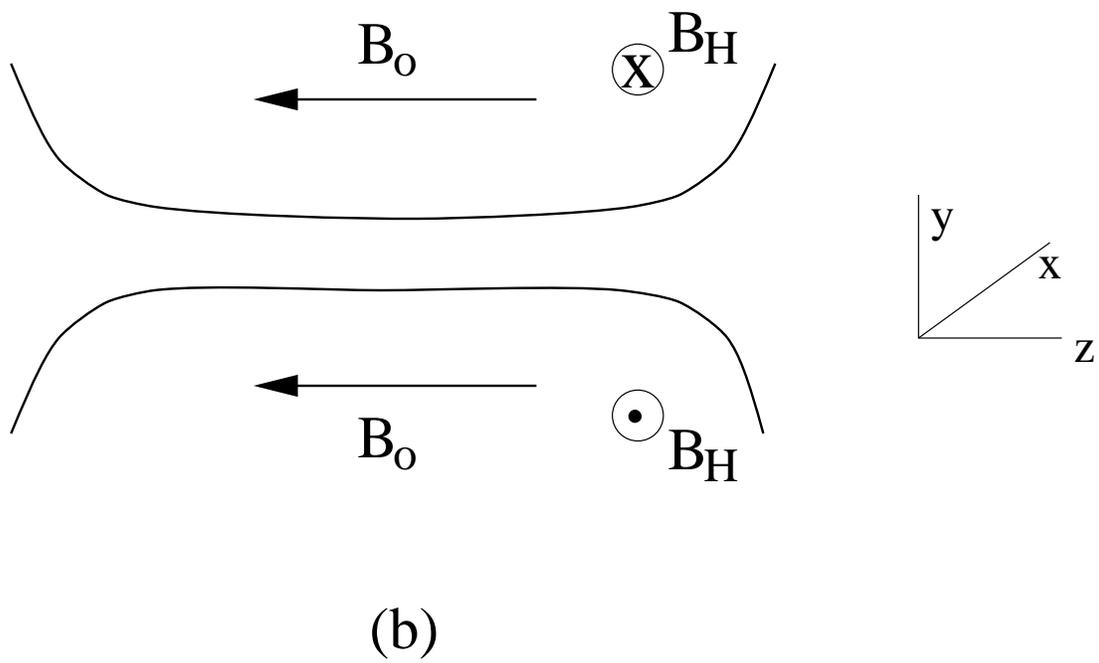
\end{document}